\documentclass[letter]{aa}  
\usepackage[varg]{txfonts}
\usepackage{natbib}
\usepackage{graphicx}
\usepackage{multicol}

\begin{document}
\title{Discovery of a magnetic field in the B pulsating system HD\,1976\thanks{Based on observations obtained at the Telescope Bernard Lyot
(USR5026) operated by the Observatoire Midi-Pyr\'en\'ees, Universit\'e de
Toulouse (Paul Sabatier), Centre National de la Recherche Scientifique of
France.}}
\titlerunning{The magnetic field of the B5IV binary HD\,1976}

\author{C. Neiner\inst{1}
\and A. Tkachenko\inst{2}\thanks{Postdoctoral Fellow of the Fund for Scientific Research (FWO), Flanders, Belgium}
\and the MiMeS collaboration
}

\offprints{C. Neiner}

\institute{LESIA, Observatoire de Paris, CNRS UMR 8109, UPMC, Universit\'e Paris Diderot, 5 place Jules Janssen, 92190 Meudon, France; 
\email{coralie.neiner@obspm.fr}
\and Instituut voor Sterrenkunde, KU Leuven, Celestijnenlaan 200D, 3001, Leuven, Belgium
}

\date{Received 7 February 2014; accepted 13 February 2014}
 
\abstract
% context heading (optional)
{}
% aims heading (mandatory)
{The presence of a magnetic field can have a strong impact on the evolution of a
binary star. However, only a dozen of magnetic OB binaries are known as of today
and available to study this effect, including very few magnetic pulsating
spectroscopic OB binaries. We aim at checking for the presence of a magnetic field in the B5IV hierarchical triple system \object{HD\,1976} with spectropolarimetric data obtained with
Narval at the Bernard Lyot Telescope (TBL).}
% methods heading (mandatory)
{We use orbital parameters of \object{HD\,1976} available in the literature to
disentangle the Narval intensity spectra. We compute Stokes V profiles with the
Least Square Deconvolution (LSD) technique to search for magnetic signatures. We
then derive an estimate of the longitudinal magnetic field strength for each observation and for various line lists.}
% results heading (mandatory)
{Our disentangling of the intensity spectra shows that \object{HD\,1976} is a
double-lined spectroscopic (SB2) binary, with
the lines of the secondary component about twice broader than the ones of the
primary component. We do not identify the third component. Moreover, we find that clear magnetic signatures are present
in the spectropolarimetric measurements of HD\,1976 and seem to be associated
with the primary component. We conclude that \object{HD\,1976} is a magnetic slowly-pulsating double-lined
spectroscopic binary star, with an undetected third component. It is the second such example known (with \object{HD\,25558}).}
% conclusions heading (optional)
{}


\maketitle
%________________________________________________________________

%\linenumbers
\section{Introduction}\label{intro}

The magnetic field of OB binaries has not been much explored so far, from either
an observational or theoretical point-of-view, while the presence of a magnetic
field in a binary could have an important impact on the evolution of the binary.

In low-mass stars, tidal interactions are expected to induce large-scale 3D
shear and/or helical flows in stellar interiors that can significantly perturb
their magnetic dynamo. Similar flows may also influence the fossil magnetic
fields of high-mass stars. Moreover, in combination with tides, magnetically
driven winds/outflows in binary systems have long been suspected to be
responsible for their orbital evolution, while magnetospheric interactions have
been proposed to enhance stellar activity. In addition, the incidence of
magnetic stars among OB binary systems provides a basic constraint on the
detailed fossil origin of the magnetic field and whether
such magnetic fields suppress binary formation. However, the crucial
observational constraints required to study these various issues are, at
present, nearly nonexistent.

In the frame of the MiMeS (Magnetism in Massive Stars) project
\citep[e.g.][]{neiner2011,wade2013}, we checked for the presence of a magnetic
field in HD\,1976.

\object{HD\,1976} (V746\,Cas) is a bright (V=5.6), slowly pulsating B (SPB) star
\citep{decat2007,dubath2011} of spectral type B5IV. Its main period of variation
is 1.065 d \citep[Hipparcos,][]{perryman1997}. \cite{decat2007}
and \cite{mathias2001} found at least one additional period at 2.5 d and
proposed  additional candidate periods. \object{HD\,1976} seems to be a
young star of $\sim$60 Myr \citep{decat2007,tetzlaff2011}, but note that
photometric age determination can lead to uncertainties for multiple systems.

\object{HD\,1976} is known to be a hierarchical triple system consisting
of a visual binary with a magnitude difference of about 0.9 magnitude
\citep{perryman1997, mason2009}, one of the component being itself a SB1 binary
\citep{decat2007}. The angular separation of the visual components
is 0.157 arcsec according to Hipparcos \citep{perryman1997} and 0.111 arcsec
according to \cite{mason2009}. The period of the visual binary is about 150
years \citep{abt2005}. The SB9 catalogue \citep{pourbaix2004} provides a period
for the SB1 between 25.4 and 27.8 d with an eccentricity between 0.12 and 0.2.

In this paper we report on Narval spectropolarimetric observations of HD\,1976
(Sect.~\ref{obs}). We use orbital parameters available in the literature to
disentangle the spectra of the two components (Sect.~\ref{binary}). We
then analyse the magnetic measurements with the Least Square Deconvolution (LSD)
technique (Sect.~\ref{analysis}) and discuss the results (Sect.~\ref{discuss}).

\section{Narval observations}\label{obs}

%%%%%%%%%%%%%%%%%%%%%%%%%%%%%%%%%%%%%%%%%%%%%%%%%%%%%%%%%%%%%%
\begin{table}[!ht]
\caption[]{Journal of 16 Narval observations of HD\,1976.}
\begin{center}
\begin{tabular}{llllll}
\hline
\hline
\# & Date & mid-HJD    & T$_{\rm exp}$ & S/N & Status  \\  
   & 2012 & $-$2456000 & s             &     &         \\
\hline
1   & 12aug12 & 152.64640 & 4$\times$900  &  560 &  good  \\
2   & 13aug12 & 153.57493 & 4$\times$900  & 1010 &  good  \\
3   & 15aug12 & 155.58735 & 4$\times$900  &  620 &  good  \\
4   & 04sep12 & 175.53081 & 4$\times$1200 & 1170 &  problem  \\
5   & 05sep12 & 176.56314 & 4$\times$1200 & 1260 &  problem?  \\
6   & 06sep12 & 177.55047 & 4$\times$1200 & 1280 &  problem?  \\
7   & 07sep12 & 178.55895 & 4$\times$1200 & 1010 &  problem  \\
8   & 08sep12 & 179.55751 & 4$\times$1200 & 1040 &  problem?  \\
9   & 11sep12 & 182.58793 & 4$\times$1200 & 1050 &  problem  \\
10  & 17sep12 & 188.48424 & 4$\times$1200 &  260 &  doubtful  \\
11  & 19sep12 & 190.50863 & 4$\times$1200 & 1260 &  doubtful  \\
12  & 01oct12 & 202.48227 & 4$\times$1200 & 1260 &  good  \\
13  & 02oct12 & 203.47914 & 4$\times$1200 & 1360 &  good  \\
14  & 03oct12 & 204.48718 & 4$\times$1200 &  830 &  good  \\
15  & 12oct12 & 213.51685 & 4$\times$1200 &  760 &  good  \\
16  & 13oct12 & 214.39089 & 4$\times$1200 & 1150 &  good  \\
\hline		
\end{tabular}		
\tablefoot{The S/N given in col. 5 is the one measured around 5000 \AA\ in the I
spectrum. Column 6 indicates the status of the Stokes V measurement related to a
technical problem (see Sect.~\ref{obs}).}	
\end{center}		
\label{lognarval}	
\end{table}		
%%%%%%%%%%%%%%%%%%%%%%%%%%%%%%%%%%%%%%%%%%%%%%%%%%%%%%%%%%%%%%

Narval is a fibre-fed echelle spectropolarimeter with a resolving power of 65000
installed at the Bernard Lyot Telescope (TBL) in France. The spectrograph covers
the wavelength domain from 3694 to 10483 \AA. We collected 16 Stokes V
measurements of \object{HD\,1976} with Narval between August and October 2012.
Their signal-to-noise ratio (S/N) varies between 260 and 1360 around 5000 \AA\
in the I spectrum (see Table~\ref{lognarval}). Since the diameter of the
Narval fiber is 2.8 arcsec, the light of all three components of
\object{HD\,1976} has been recorded in the spectra.

During some of our observations, a loss of reference of one of the rhomboedra
occurred: the coder could not read the exact position of the rhomboedra
due to a small wobble of the disk. The lack of reading was compensated by the
closed-loop system by changing the speed of the motor hence making the
rhomboedra jump to an unknown angle (with the coder position kept
constant) until the reading resumed. This technical problem does not affect the
intensity spectra. It does however affect the Stokes V
measurements, as the angle of the rhomboedra was not the one needed to measure
the full circular polarisation. When this problem occurred, the Stokes V
signature has not been increased, but it could have been decreased or its shape
modified, as some of the circularly polarised light was lost. In other words,
this problem cannot lead to the appearance of spurious signatures, but it can
modify the signatures.

The spectra obtained in August and October are not affected by this technical
problem. It only occured in September. By checking known magnetic stars
obtained during the same nights as our observations of \object{HD\,1976}, we
found that the spectra obtained on September 4, 7 and 11 are probably affected
by this technical issue. For those obtained on September 5, 6 and 8, there is no
magnetic star available to check the reliability of the Stokes V measurements
but, since they were obtained in the same week as the others, we consider them
probably affected as well. The two spectra obtained on September 17 and 19 are
part of a different observing sequence, but since we have no magnetic check star
for these observations either, we consider them doubtful. The status of each
spectrum is reported in Table~\ref{lognarval}.

\begin{figure}[!ht]
\begin{center}
\resizebox{0.74\hsize}{!}{\includegraphics[clip]{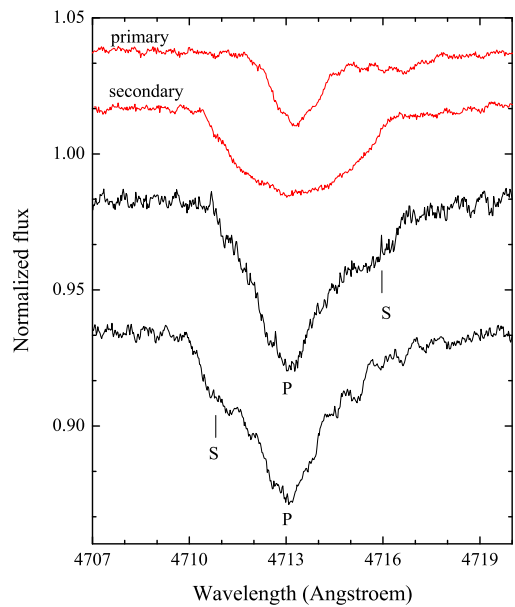}}
\caption[]{Bottom (black curves): Examples of two spectra (\#1 and \#16) of
the \ion{He}{i} line at 4713 \AA\ obtained at two different phases of the
orbital period. The two binary components are indicated (P and S). Top
(red curves): disentangled spectra of the \ion{He}{i} line at 4713 \AA\  for the
primary and secondary components. All spectra are shifted vertically to allow
for a better reading.}
\label{SB2}
\end{center}
\end{figure}

Data were reduced with the {\sc Libre-Esprit} reduction package, an extension of
{\sc Esprit} \citep{donati1997} for Narval available at the telescope. The usual
bias and flat-field corrections were applied, as well as a wavelength
calibration with a ThAr lamp. Each echelle order was then carefully normalised
with the continuum package of the {\sc IRAF} software\footnote{IRAF is
distributed by the NOAO, which is operated by the AURA under cooperative
agreement with the NSF.}. 		

\section{Possible system configurations of HD\,1976}\label{binary}

Characteristic asymmetries are detected in several individual lines of He and Mg
in the spectra of \object{HD\,1976} (see Fig.~\ref{SB2}), suggesting a
double-lined nature of HD\,1976. Similar asymmetries are also seen in the LSD
Stokes I profiles (see  Sect.~\ref{analysis} below).

To verify whether the signal could be attributed to the close companion star
of the SB1 or to the visual binary, we applied the spectral disentangling
({\sc spd}) technique to our 16 spectra. For this purpose, we used the {\sc
FDBinary} code \citep{ilijic2004}, which relies on the Fourier-based
implementation of the {\sc spd} method suggested by \citet{hadrava1995}.
Compared to the original method proposed by \citet{simon1994} that applies to
the spectra in wavelength domain, the Fourier-based {\sc spd} technique has the
clear advantage of being much faster. This makes the method applicable to
time-series of high-resolution spectroscopic data.

\begin{figure*}[!ht]
\begin{center}
\resizebox{\hsize}{!}{\includegraphics[clip]{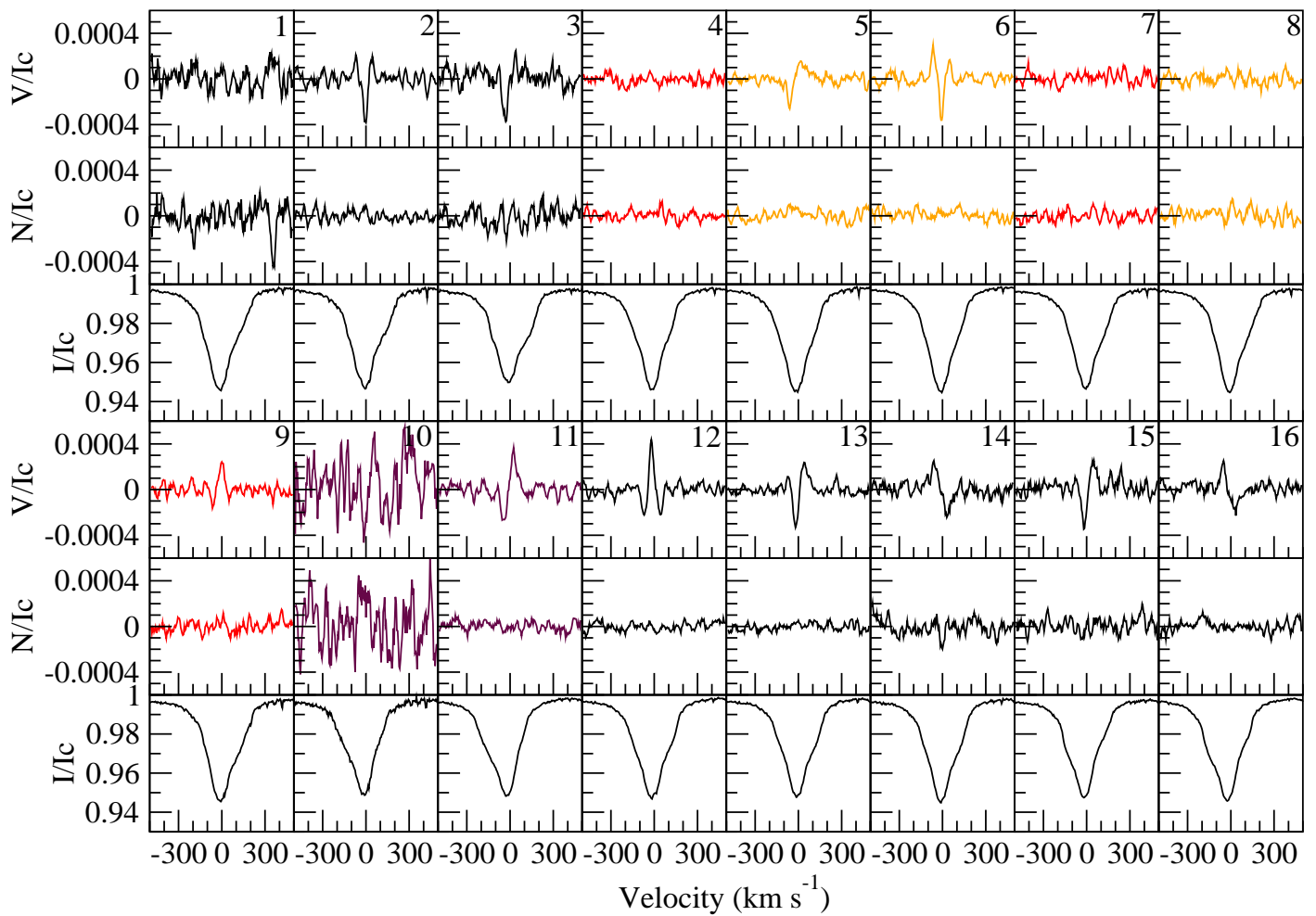}}
\caption[]{LSD Stokes V, N and I profiles computed for the composite spectra of
HD\,1976 with the mask containing all lines. The spectra affected by a technical
problem are shown in red, those that are probably affected in orange, those that
are doubful in maroon, those that did not undergo a problem are in black. The
reference number of the spectrum is indicated in the corner of each Stokes V
panel.}
\label{compositeBl}
\end{center}
\end{figure*}

No precise orbital solution is available for \object{HD\,1976} in the
literature. The poor phase coverage and limited number of measurements we have
at our disposal do not allow us to determine orbital elements either. Therefore,
we used all three orbital periods available in the SB9 catalogue
\citep{pourbaix2004}: 25.4176, 25.44, and 27.8 d. We ran the {\sc FDBinary} code
in pure separation mode (without optimizing for the light factors) focusing on
wavelength regions centered at several strong \ion{He}{i} (5016 and 4713 \AA),
\ion{Mg}{ii} (4481 \AA) and \ion{C}{ii} (4267 \AA) lines.

We obtained consistent results for all considered spectral regions and values of
the orbital period, i.e. spectral contribution of two stars 
could be detected in all cases. Fig.~\ref{SB2} illustrates two observed 
composite spectra (\#1 and \#16) and the disentangled spectra of the primary and
secondary components. The line profile of the secondary component spans about
374 km~s$^{-1}$ and varies significantly in radial velocity, while the
profile of the primary spans only about 184 km~s$^{-1}$ and is
relatively stable in radial velocity. Therefore, the secondary component is
clearly associated to the SB1, and the primary component could be either the
other component of the SB1 (with a high mass ratio between the  two stars) or
the visual component or the sum of both.

We conclude that \object{HD\,1976} is a double-lined spectroscopic binary (SB2)
star with a spectroscopically undetected third component, but more data
is required to obtain a precise orbital solution and identification of
the various components in the spectra, as well as the disentangled spectra
suitable for detailed spectrum and magnetic analysis of the individual
components.

\section{Spectropolarimetric analysis}\label{analysis}

We applied the LSD technique \citep{donati1997} to the Narval data. To this aim,
we first constructed a spectral line list from the VALD database
\citep{piskunov1995, kupka1999} with T$_{\rm eff}=16000$ K and $\log g=4.0$ dex
\citep{mathias2001}. The list only contains lines with a depth above 1\% of the
continuum level. From this line template we removed all hydrogen lines, lines
that are blended with H lines or interstellar bands, as well as lines that did
not seem to be present in the Narval spectra. We then adjusted the strength of
the remaining lines to fit the Narval observations. This resulted in a mask
containing 121 lines. From this initial mask, we then also created two submasks
containing only the 18 unblended He lines or 85 metal lines except He and lines
blended with He. 

Using these three line masks, we extracted LSD Stokes I and V profiles for each
spectropolarimetric measurements. We also extracted null (N) polarisation
profiles to check for spurious signatures, e.g. from instrumental origin or 
stellar pulsations. For the mask with only He lines the mean wavelength is 4832
\AA\ and the mean Land\'e factor is 1.127, for the mask without He lines these
parameters are 5029 \AA\ and 1.152, for the mask with all lines they are 4985
\AA\ and 1.196.

%%%%%%%%%%%%%%%%%%%%%%%%%%%%%%%%%%%%%%%%%%%%%%%%%%%%%%%%%%%%%%
\begin{table}[!ht]
\caption[]{Longitudinal magnetic field measurements of HD\,1976 measured in the composite spectra with three different line masks.}
\begin{center}
\begin{tabular}{lllllll}
\hline
\hline
\# & \multicolumn{2}{l}{All} & \multicolumn{2}{l}{He lines} & \multicolumn{2}{l}{All but He} \\  
   & $B_l$ & $\sigma_{B_l}$ & $B_l$ & $\sigma_{B_l}$ & $B_l$ & $\sigma_{B_l}$ \\  
   & G     & G & G     & G & G     & G \\
\hline
1   &	4.4  &  75.1  &  191.6 & 106.8 &  -34.0 & 267.7\\
2   &  22.8  &  33.2  &   44.4 &  46.0 &  -67.9 & 106.1\\
3   &-149.2  &  54.8  & -164.6 &  78.5 & -198.1 & 169.7\\
(4) & 7.1    &  28.9  &  -24.4 &  41.4 &   81.6 &  80.7\\
(5) &-153.5  &  25.3  & -111.5 &  35.5 & -407.1 &  83.0\\
(6) &  56.7  &  25.2  &   41.8 &  35.0 &   88.5 &  77.4\\
(7) &  11.2  &  33.3  &  -55.9 &  46.9 &   44.1 &  97.5\\
(8) & -12.5  &  31.4  &  -58.9 &  43.8 &  137.9 & 101.0\\
(9) & 3.5    &  31.0  &	  46.1 &  43.5 &  -69.4 & 113.0\\
(10)& -91.7  & 163.3  & -213.8 & 262.1 &  100.8 & 410.5\\
(11)&-208.1  &  27.6  & -222.3 &  39.1 & -292.4 &  74.7\\
12  &	5.6  &  26.5  &  -65.1 &  36.7 &   50.5 &  85.0\\
13  & -83.0  &  25.5  &  -11.4 &  35.4 & -296.3 &  72.1\\
14  & 206.4  &  39.6  &  132.2 &  55.2 &  554.7 & 127.6\\
15  &-107.1  &  44.7  & -111.9 &  63.8 & -153.3 & 123.7\\
16  & 156.1  &  28.5  &  134.3 &  39.7 &  352.5 &  91.7\\
\hline		
\end{tabular}		
\tablefoot{Measurements from \#5 to \#11, indicated in brackets, should be
considered with care due to a technical issue with the polarimeter (see
Sect.~\ref{obs} and Table~\ref{lognarval}).}	
\end{center}		
\label{bl}	
\end{table}		
%%%%%%%%%%%%%%%%%%%%%%%%%%%%%%%%%%%%%%%%%%%%%%%%%%%%%%%%%%%%%%

We find that the LSD Stokes V profiles show clear signatures of a magnetic
field. In particular all Stokes V profiles obtained during good conditions (good
S/N and no technical problem) show a clear magnetic signature (see the first and
fourth rows of Fig.~\ref{compositeBl}). On the contrary, the N profiles do not
show signatures, which confirms that the magnetic measurements have not been
contamined by the stellar pulsations, binary or instrumental polarisation
effects (see the second and fifth rows of Fig.~\ref{compositeBl}). 

The magnetic signatures are observed over a velocity range that spans about 200
km~s$^{-1}$ and remain stable in radial velocity, therefore
they seem to be associated with the primary component for which the line
profiles also span this velocity range and which does not vary much in radial
velocity. However, we cannot exclude that the secondary component is also
magnetic (with a weaker field) without an accurate disentangling of the spectra.

The longitudinal field ($B_l$) values and their error bars ($\sigma_{B_l}$),
extracted from the LSD Stokes V profiles in the range from -120 to 120
km~s$^{-1}$ using each mask, are reported in Table~\ref{bl}. However, since
these values are extracted with the composite I spectrum of the two components
of \object{HD\,1976}, they do not represent the real field values of the
magnetic component but an underestimate. We also recall that the values
extracted certain nights might be affected by a technical issue with the
rhomboedra position (see Sect.~\ref{obs}). Those $B_l$ values should thus be
considered with care and are marked in Table~\ref{bl}.

\section{Discussion}\label{discuss}

%\subsection{The two components of HD\,1976}

Using orbital parameters available in the SB9 catalogue \citep{pourbaix2004}, we
have been able to disentangle the spectra of the two components of
\object{HD\,1976}. This shows that HD\,1976 is a SB2 binary. The primary
component has a lower $v\sin i$ value ($\sim$90 km~s$^{-1}$) than the secondary
component ($\sim$190 km~s$^{-1}$). The third component remains hidden in
our spectra.

However, various orbital periods are proposed in the literature and we
could not distinguish between them, nor identify precisely the origin of
the primary component. Therefore our disentangled spectra are too uncertain to
be used to compute reliable magnetic measurements of the individual components.

%\subsection{The magnetic field of the primary star of HD\,1976}

Nevertheless, the Narval spectropolarimetric measurements clearly show
that \object{HD\,1976} is a magnetic star. The magnetic signatures seem to be
associated with the primary component. 

The longitudinal magnetic field measured using all lines varies between about
-200 and 200 G. These values are however underestimated for two reasons: first
we used the composite intensity spectra to derive them and thus the Stokes V
spectra were normalized with a too strong intensity, second a technical problem
with the polarimeter rhomboedra led to a loss of part of the circular
polarisation signal for some of the measurements.

Assuming an oblique dipole field, as observed in most massive stars and as
suggested by the shape of the observed Stokes V profiles (see
Fig.~\ref{compositeBl}), and following \cite{schwarzschild1950}, we estimate the
polar field strength to be of at least 600 G according to the above
longitudinal field values. Considering that these values are underestimated, the
polar field might be much stronger.

\section{Conclusions}

\object{HD\,1976} is a very interesting object as known magnetic pulsating
spectroscopic binary are rather rare. The only other known SPB example is
\object{HD\,25558} \citep{sodor2014}.

We plan on acquiring new spectra of \object{HD\,1976} to derive a more precise
orbital solution. We will then be able to disentangle the spectra more reliably
and measure the magnetic field of the individual components. This is necessary
to check whether the secondary component is also magnetic as well as to derive
more accurate measurements of the magnetic field of the primary. Moreover,
we will be able to test whether the magnetic component is part of the SB1 and is
the pulsating component.

Finally, the BinaMIcS (Binarity and Magnetic Interactions in various
classes of stars) project \citep[see][]{neiner2013}, which relies on two large
programs of observations with ESPaDOnS at CFHT and Narval at TBL, will allow us
to acquire spectropolarimetric observations of many magnetic binary systems of
all spectral types. These new observations will provide additional examples
of magnetic massive binary stars, in particular systems with short
orbital periods. Comparing HD\,1976 with other magnetic massive binaries will be
very useful to derive information about the role of magnetic field on binarity
and vice-versa. 

\begin{acknowledgements}
CN wishes to thank the Programme National de Physique Stellaire (PNPS) for their
support. The research leading to these results has partly received funding from
the Fund for Scientific Research of Flanders (FWO), Belgium, under grant
agreement G.0B69.13. This research has made use of the SIMBAD database operated
at CDS, Strasbourg (France), and of NASA's Astrophysics Data System (ADS).
\end{acknowledgements}

\bibliographystyle{aa}
\bibliography{articles}

\end{document}